\documentclass[UTF-8,preprintnumbers,superscriptaddress,aps,reprint,showkeys]{revtex4}
\usepackage{amsmath,amssymb,graphicx,bm}

\begin{document}
	\title{Possible signatures of higher dimension in thin accretion disk around brane world black hole}

\author{Ailin Liu}
	\affiliation{College of Physics Science and Technology, Hebei University, Baoding 071002, China}	

\author{Tong-Yu He}
\affiliation{College of Physics Science and Technology, Hebei University, Baoding 071002, China}

\author{Ming Liu}
\affiliation{College of Physics Science and Technology, Hebei University, Baoding 071002, China}

\author{Zhan-Wen Han}
\affiliation{College of Physics Science and Technology, Hebei University, Baoding 071002, China}
\affiliation{Yunnan Observatories, Chinese Academy of Sciences, Kunming 650216, China}

\author{Rong-Jia Yang \footnote{Corresponding author}}
\email{yangrongjia@tsinghua.org.cn}
\affiliation{College of Physics Science and Technology, Hebei University, Baoding 071002, China}
\affiliation{Hebei Key Lab of Optic-Electronic Information and Materials, Hebei University, Baoding 071002, China}
\affiliation{National-Local Joint Engineering Laboratory of New Energy Photoelectric Devices, Hebei University, Baoding 071002, China}
\affiliation{Key Laboratory of High-pricision Computation and Application of Quantum Field Theory of Hebei Province, Hebei University, Baoding 071002, China}

\begin{abstract}
 We probe deeply into the characteristics of thin accretion disk surrounding black hole within the brane world paradigm. We investigate how model parameters affect the physical properties of the disk. Our findings indicate that as the tidal charge parameter inherited from the higher dimension increases, the energy flux, the radiation temperature, the spectral cutoff frequency, the spectral luminosity, and the conversion efficiency of the disk all increase, but the radius of the innermost stable circular orbit decreases. Compared to cases of the Kerr and Schwarzschild black holes, the disk is hotter and more luminous for positive tidal charge parameter, while it is cooler and less luminous for negative tidal charge parameter, which suggests the potential for probing possible signatures of higher dimension.
\end{abstract}
%	\pacs{}
\maketitle

\section{Introduction}
Black holes (BHs) have long been a focal point of research in both general relativity (GR) and higher-dimensional gravity theories. The brane world gravity has sparked new research interests in this field. The brane world theory is an extension of gravitational theory, proposing that our universe may be a brane or hypersurface embedded in a higher-dimensional spacetime, the physical fields (electro-magnetic, Yang-Mills etc.) are confined to the three brane, only gravity can
freely propagate in both the brane and bulk space-times, with the gravitational self-couplings not significantly modified \cite{Randall:1999ee, Randall:1999vf}. These additional dimensions exist beyond the three spatial dimensions we typically observe and can only be detected through specific physical phenomena or experiments \cite{Arkani-Hamed:1998jmv, Antoniadis:1998ig,Yu:2016tar,Lin:2023zna}. In the development of this scenario, two brane models have been proposed: the thin brane model and the thick brane model. The well-known thin brane models, such as the Arkani-Hamed-Dimopoulos-Dvali (ADD) thin brane model and the Randall-Sundrum (RS) thin brane model, were primarily developed to address the hierarchy problem \cite{Arkani-Hamed:1998jmv,Antoniadis:1998ig,Randall:1999ee}. In \cite{Dadhich:2000am}, a detailed discussion and exposition of the exact static spherically symmetric vacuum solutions in the brane world model was provided for the first time. Recently, a general algorithm that enables the consistent embedding of any four-dimensional static and spherically symmetric geometry into any five-dimensional single-brane braneworld model was proposed \cite{Nakas:2023yhj}. In \cite{Bronnikov:2003gx,Aliev_2005}, the authors explored the precise steady-state solutions and axisymmetric solutions of BHs, validating the formation of a closed system in the effective gravitational field equations on the brane with the Kerr-Schild form of the metric, replacing the Kerr-Newman solutions from ordinary GR with tidal charge replacing the electric charge. The charged rotating BH (hereafter for short: BWBH) obtained in \cite{Aliev_2005} has been studied extensively, see for example \cite{Pun:2008ua,Amarilla:2011fx,Banerjee:2019sae,Tsukamoto:2017fxq,Vagnozzi:2019apd,Nucamendi:2019qsn,Hou:2021okc,
Mishra:2021waw,Shahzadi:2022rzq,Wei:2022jbi,Vagnozzi:2022moj,Roy:2023rjk,Zi:2024dpi,Stuchlik:2008fy,Bohra:2023vls}. These studies involved exploring the nature and effects of extra dimensions and their connection and divergence from traditional four-dimensional gravitational theories. Here we will focus on the study of thin accretion disks around such a BWBH and probe possible signatures of higher dimensions in relative physical quantities.

The first comprehensive theory regarding accretion disks around BHs was established in \cite{Shakura:1972te}, which was later extended to the mass accretion of a rotating BH in GR \cite{Novikov:1973kta}. Further discussions on the characteristics of radiation energy flux on thin accretion disks are elaborated in \cite{Page:1974he} and \cite{Thorne:1974ve}. Recently, Novikov-Thorne model has been applied to test various BHs, see for example \cite{Zhu:2019ura,Heydari-Fard:2021ljh,Zhang:2021hit,Karimov:2018whx,Kazempour:2022asl,Gyulchev:2021dvt,Collodel:2021gxu, He:2022lrc,Feng:2024iqj,Shahidi:2020bla,Uniyal:2023inx,Uniyal:2022vdu,Staykov:2016dzn, Teodoro:2021ezj, Rahman:2022fay}. These studies provide us with a more comprehensive model of BH behavior and contribute to a deeper understanding of the interaction and energy conversion mechanisms between BHs and accretion disks. Physical properties of thin disk around BWBHs were also investigated in \cite{Pun:2008ua, Banerjee:2019sae}, but the formula of the flux emanating from the accretion disk in their works is incorrect, and the related results obtained from it are also unreliable. So, it is necessary to reconsider this issue.

The structure of this paper is outlined as follows. In Sect. II, we will provide a concise overview of the BWBHs, followed by presenting the geodesic equations in the equatorial plane. Sect. III will delve into an exploration of the physical attributes of thin accretion disks surrounding BWBHs, focusing on the impact of tidal charge parameters on the energy flux, the radiation temperature, and the emission spectrum. Finally, a brief summary will be presented in the fourth section.

\section{Brane world black holes and stable circular orbits}
In this section, we will briefly review the solutions of BWBHs and give the geodesic equations governing the motion of particles within the equatorial plane.

\subsection{Brane world black holes}
The Kerr-Schild form of metric is a mathematical assumption used to describe rotating BHs. This metric form offers certain simplifications and conveniences in mathematics, making it easier to study the spacetime structure around BHs. In the RS brane world theory, the Kerr-Schild form of metric was employed to describe rotating BHs localized on 3-branes \cite{Dadhich_2000}. In \cite{Aliev_2005}, the authors used this metric form and introduced an additional tidal charge parameter $\beta$ to account for gravitational effects from extra dimensions. They derived the various components of the Kerr-Schild metric, including $g_{tt}$, $g_{t \varphi}$, $g_{rr}$, $g_{\theta\theta}$ and $g_{\varphi\varphi}$. Subsequently, by applying the Boyer-Lindquist transformation, they obtained the metric expression describing rotating BHs. The metric expression are given as follows:
\begin{eqnarray}
\label{1}
 \mathrm{d}s^{2}&=&-g_{tt} \mathrm{d}t^{2}+2g_{t\phi}\mathrm{d}t \mathrm{d}\phi+g_{rr} \mathrm{d}r^{2} +g_{\theta \theta } \mathrm{d}\theta  ^{2} +g_{\phi \phi }\mathrm{d}\phi  ^{2}\\\nonumber
&=&-\left(1-\dfrac{2Mr-\beta}{\Sigma}\right)\mathrm{d}t^{2}-\dfrac{2a\left(2Mr-\beta\right)}{\Sigma}\sin^{2}\theta	\mathrm{d}t	\mathrm{d}\phi\\\nonumber
&+&\dfrac{\Sigma}{\Delta}\mathrm{d}r^{2}+\Sigma\mathrm{d}\theta^{2}+\left(r^{2}+a^{2}+\dfrac{2Mr-\beta}{\Sigma}a^{2}\sin^{2}\theta\right)\sin^{2}\theta\mathrm{d}\phi^{2},
\end{eqnarray}
where
\begin{equation}
\label{3}
\Sigma=r^{2}+a^{2}\cos^{2}\theta,
\end{equation}
\begin{equation}
\label{4}
\Delta=r^{2}+a^{2}-2Mr+\beta.
\end{equation}
The symbol $M$ represents the mass of the BH, $a$ denotes the angular momentum, and $\beta$ stands for the tidal charge parameter which is inherited from the higher dimensions and can be both positive and negative values. For positive $\beta$, Eq. \eqref{1} is similar to a Kerr-Newmann BH, while the case for negative $\beta$ has no analogue in GR and thus provides a true signature of the additional spatial dimensions. The ADM mass of BWBH is $M$. When $\beta =0$, the BH is simplified to the Kerr BH in GR. The event horizon is determined by the equation $\Delta=0$, resulting
\begin{equation}\label{5}
r_{\pm}=M\pm\sqrt{M^{2}-a^{2}-\beta}.
\end{equation}
$r_{+}$ denoted as the largest root, representing the location of the outer event horizon. The structure of this event horizon is contingent upon the sign of the tidal charge. For the event horizon to exist, the condition $M^{2}\geqslant a^{2}+\beta$ must be satisfied, where equality corresponds to the limiting horizon.

\subsection{Stable circular orbits}
The timelike geodesics around compact astrophysical objects are very significant for theoretical as well as experimental point of
view as they convey the information about the compact objects where very strong gravity takes place. Considering the geodesic of radial timelike particles in the equatorial plane ($\theta=\pi/2$) in which the accretion disk is usually assumed to lie, the Lagrangian takes the following form
\begin{eqnarray}
\mathfrak{L}=-g_{t t} \dot{t}^2+2 g_{t \phi} \dot{t} \dot{\phi}+g_{r r} \dot{r}^2+g_{\phi \phi} \dot{\phi}^2,
\end{eqnarray}
where the dot means the derivative with respect to the affine parameter $\lambda$ of the world line of the particle. From Euler-Lagrangian equation, we have
\begin{equation}
\begin{aligned}
-g_{t t} \dot{t}+g_{t \phi} \dot{\phi} & =-E, \\
g_{t \phi} \dot{t}+g_{\phi \phi} \dot{\phi} & =L,
\end{aligned}
\end{equation}
where $E$ and $L$ are two constants of motion for particles which are known as the specific energy and angular momentum of the particles moving on circular orbits around the compact objects. Solving the above two equations, we derive the orbiting particle's four-velocity components as
\begin{equation}\label{7}
 \dot{t}=\frac{Eg_{\phi \phi }+Lg_{t\phi }  }{g_{t\phi } ^{2}-g_{tt}g_{\phi \phi } },
\end{equation}
\begin{equation}\label{8}
\dot{ \phi}=-\frac{Eg_{t\phi }+Lg_{tt}  }{g_{t\phi }^{2} -g_{tt} g_{\phi \phi }  }.
\end{equation}
The first integral of geodesics equation ($g_{\mu\nu}\dot{x}^{\mu}\dot{x}^{\nu}=-1$) for time-like particles, yields
\begin{equation}
g_{r r} \dot{r}^2=-1+g_{t t} \dot{t}^2-2 g_{t \phi} \dot{t} \dot{\phi}-g_{\phi \phi} \dot{\phi}^2 .
\end{equation}
Putting the values of $\dot{t}$ and $\dot{ \phi}$, we have
\begin{equation}
g_{r r}\left(\frac{d r}{d \tau}\right)^2=-1+\frac{E^2 g_{\phi \phi}+2 E L g_{t \phi}+L^2 g_{t t}}{g_{t \phi}^2+g_{t t} g_{\phi \phi}},
\end{equation}
from which we can define an effective potential as
\begin{equation}\label{10}
	V_{\text{eff}} =-1+\frac{E^{2}g_{\phi \phi }+2ELg_{t\phi }+L^{2} g_{tt} }{g_{t\phi }^{2}-g_{tt}g_{\phi \phi } }.
\end{equation}

Circular orbits in the equatorial plane are satisfied the conditions $V_{\text{eff}} = 0$, $V_{\text{eff},\text{r}} = 0$, and $V_{\text{eff},\theta} = 0$ \cite{Gair_2008,Bambi:2011jq,Bambi:2011vc}. Due to the reflection symmetry of the metric about the equatorial plane, particles located at $\theta=\frac{\pi}{2}$ naturally fulfill the condition $V_{\text{eff},\theta} = 0$. Utilizing these conditions, we derived the expressions for the specific energy $E$, the specific angular momentum $L$, and the specific angular velocity $\Omega$ for particles undergoing prototype orbital motion in the equatorial plane of a rotating Kerr-like BH spacetime.
\begin{equation}\label{11}
	\Omega =\dfrac{\mathrm{d}\phi}{\mathrm{d}t}=\frac{-g_{t\phi,r } +\sqrt{(g_{t\phi ,r} )^{2}  -g_{tt,r}g _{\phi \phi ,r}  }}{g_{\phi \phi ,r} },
\end{equation}
\begin{equation}\label{12}
	E=-\frac{g_{tt}+g_{t\phi }\Omega   }{\sqrt{-g_{tt}-2g_{t\phi }\Omega   -g_{\phi \phi } \Omega ^{2} } },
\end{equation}
\begin{equation}\label{13}
	L=\frac{g_{t\phi }+g_{\phi \phi } \Omega  }{\sqrt{-g_{tt}-2g_{t\phi } \Omega  -g_{\phi \phi } \Omega ^{2} } }.
\end{equation}

For test particles in the gravitational potential of a central celestial body, the innermost stable circular orbit (ISCO) is defined by the radius at which $V_{\text{eff},rr}=0$. By satisfying this condition $V_{\text{eff},rr}=0$, we can determine the marginally stable orbit around the central celestial body. We can express the effective potential as, $V_{\text{eff}}(r)\equiv-1+f/g$, where $f\equiv E^{2}g_{\phi \phi }+2ELg_{t\phi }+L^{2} g_{tt}$ and $g\equiv g_{t\phi }^{2}-g_{tt}g_{\phi \phi }$. In addition, the constraint $g \neq 0$ is imposed. By setting $V_{\text{eff}}(r) = 0$, we can obtain $f = g$. The condition $V_{\text{eff},r}(r) = 0$ yields $f,_{r}g - fg,_{r} = 0$. Consequently, with these constraints, we can deduce $V_{\text{eff},rr}(r) = 0$, which provides the following important relationship
\begin{equation}\label{17}
0=(g_{t\phi}^{2}-g_{tt}g_{\phi})V_{\text{eff},rr}
=E^{2}g_{\phi\phi},_{rr}+2ELg_{t\phi},_{rr}+L^{2}g_{tt},_{rr}-(g_{t\phi}^{2}-g_{tt}g_{\phi\phi}),_{rr}.
\end{equation}
By substituting equations (\ref{11}-\ref{13}) into equation (\ref{17}), we can solve for $r$ and obtain the radius of the edge stable orbit. Additionally, we can also determine the edge stable orbit by investigating the condition of $\mathrm{d}E/\mathrm{d}r=0$, which is exactly the method used here, as shown in the following equation:
\begin{equation}\label{18}
Mr(6Mr-r^{2}-9\beta+3a^{2})+4\beta(\beta-a^{2})-8a(Mr-\beta)^{3/2}=0
\end{equation}
When $a=\beta=0$, the edge stable orbital radius $r_{\text{isco}}$ is equal to $6M$ for Schwarzschild BH. For $r$ $< r_{\text{isco}}$ , the equatorial circular orbits are unstable, thereby
$r_{\text{isco}}$ defines the inner edge of the thin accretion disk.

\section{Physical properties of thin accretion disks}
In this section, we will investigate the accretion process in thin disks around BWBHs. We will discuss in detail the effect of the tidal charge parameter $\beta$ on the energy flux, the radiation temperature, the emission spectrum and the energy conversion efficiency, and so on.
	
We first briefly summarized the key physical characteristics of thin accretion disks according to the Novikov-Thorne model \cite{Novikov:1973kta,Shakura:1972te}: (1) The spacetime surrounding the central mass object is static, axisymmetric, and asymptotically flat. (2) The self-gravity of the accretion disk is negligible, implying that its mass does not influence the background metric. (3) The accretion disk is geometrically thin, with its vertical scale $h$ significantly smaller than its horizontal scale ($h \leq r$). (4) The radius of the marginally stable orbit dictates the inner edge of the disk and the particles move between the ISCO and the outer edge of the disk. (5) The accretion disk resides on the equatorial plane of the compact accreting object, orthogonal to the surface of the BH's spin. (6) It is assumed that the electromagnetic radiation emitted perpendicularly to the disk’s plane conforms to a black-body spectrum, which is optically thick and is influenced by the fluid dynamics and thermodynamic equilibrium conditions of the disk. (7) The mass accretion rate $\dot{M_{0}}$ remains constant over time due to the stable state of the accretion disk.
	
The physical constants and the representations of thin accretion disk used in this paper are as follows : $c=2.997\times 10^{10}$ cms$^{-1}$,  $M_{\odot}=1.989\times10^{33}$g,  $\dot{M}_{0}=2\times10^{-6} M_{\odot}$yr$^{-1}$,  1yr$=3.156\times10^7$s,  $\sigma_{\rm{SB}}=5.67\times 10^{-5}$erg s$^{-1}$cm$^{-2}$K$^{-4}$, $h=6.625\times 10^{-27}$ergs, $k_{\rm{B}}=1.38\times 10^{-16}$ ergK$^{-1}$ and the mass of BH $M=2\times 10^6M_{\odot}$, as used in \cite{He:2022lrc}. SgrA* can be seen as a realization of such a model, which is a supermassive BH at the center of the Milky Way with a mass of order $M = 4.1\times10^6M_{\odot}$ and with an estimated rate $\dot{M}_{0}=10^{-9} -10^{-7}M_{\odot}$yr$^{-1}$.
	
\subsection{The radiant energy flux}
Considering accretion disks surrounding BWBHs, the equation governing the conservation of rest mass, energy, and angular momentum for disk particles allows us to determine the radiant energy flux at the disk surface as \cite{Page:1974he,Novikov:1973kta}
\begin{equation}\label{19}
F(r)=-\frac{\dot{M_{0} } \Omega _{,r} }{4\pi\sqrt{-g}(E-\Omega L ) ^{2}  } \int_{r_{\rm{isco}} }^{r}(E-\Omega L)L_{,r}\mathrm{d}r.
\end{equation}
The mentioned equation is applicable solely in cylindrical coordinates. For spherical coordinates, we need to refer to \cite{Collodel:2021gxu}
\begin{eqnarray}
\label{20}
F(r)=-\frac{\dot{M_{0} } \Omega _{,r} }{4\pi\sqrt{-g/g_{\theta\theta}}(E-\Omega L ) ^{2}  } \int_{r_{\rm{isco}} }^{r}(E-\Omega L)L_{,r}\mathrm{d}r
\end{eqnarray}
where $\dot{{M_{0}}} $ is the accretion rate of mass.
%%%%%%%%%%%%%%%%%%%%%%%%
\begin{figure}
	\begin{minipage}{0.4\textwidth}
		\includegraphics[height=5cm,width=7cm]{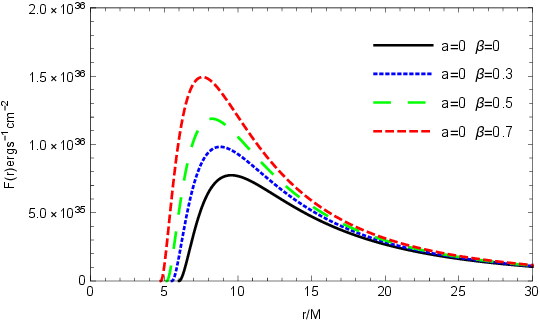}
	\end{minipage}%
	\begin{minipage}{0.4\textwidth}
		\includegraphics[height=5cm,width=7cm]{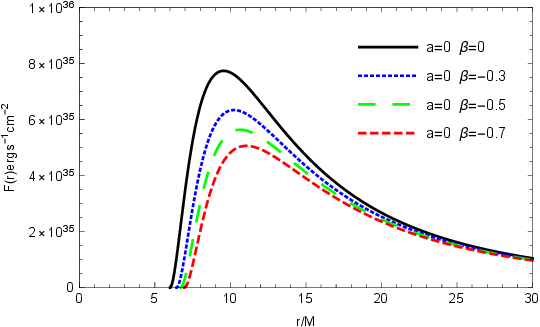}
	\end{minipage}
	\caption{The energy flux $F(r)$ of a disk around a non-rotating $(a=0)$ BWBH with different values of $\beta$.}
	\label{fig1}
\end{figure}
%%%%%%%%%%%%%%%%%%%%%%
\begin{figure}
	\begin{minipage}{0.4\textwidth}
		\includegraphics[height=5cm,width=7cm]{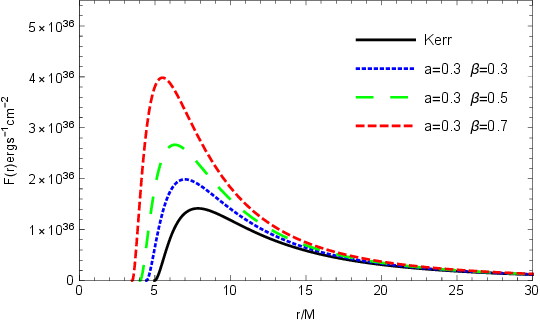}
	\end{minipage}%
	\begin{minipage}{0.4\textwidth}
		\includegraphics[height=5cm,width=7cm]{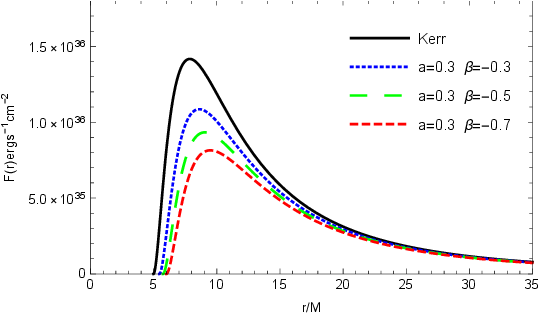}
	\end{minipage}
	\caption{The energy flux $F(r)$ of a disk around a rotating $(a=0.3)$ BWBH with different values of $\beta$.}
	\label{fig2}
\end{figure}
%%%%%%%%%%%%%%%%%%%%%%%
\begin{figure}
	\begin{minipage}{0.4\textwidth}
		\includegraphics[height=5cm,width=7cm]{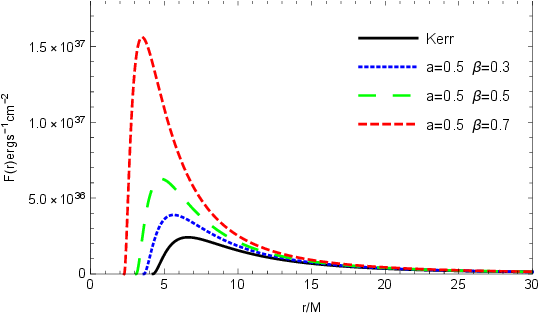}
	\end{minipage}%
	\begin{minipage}{0.4\textwidth}
		\includegraphics[height=5cm,width=7cm]{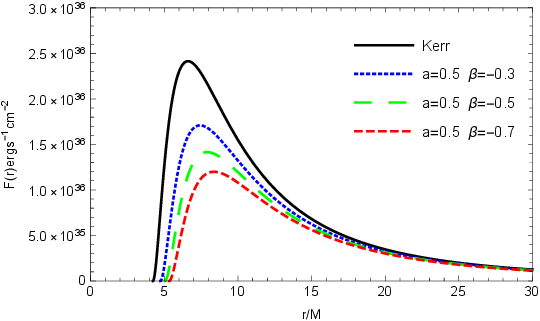}
	\end{minipage}
	\caption{The energy flux $F(r)$ of a disk around a rotating $(a=0.5)$ BWBH with different values of $\beta$. }
	\label{fig3}
\end{figure}

Figures \ref{fig1}, \ref{fig2}, and \ref{fig3} depict the energy flux on the disk surface for different values of the parameters. From figure \ref{fig1}, it is evident that as the tidal charge $\beta$ increases with a fixed rotation parameter $a$, the energy flux also increases, while the radius of the ISCO decreases. Figures \ref{fig2}, and \ref{fig3} show that as the rotation speed $a$ increases, the influence of the tidal charge $\beta$ becomes more pronounced. The energy flux values increase more rapidly as $\beta$ increases at equal intervals. On the left panel of figures \ref{fig1}, \ref{fig2}, and \ref{fig3}, it can be observed that for positive $\beta$, the energy flux exceeds that of the Kerr BH ($a\neq0$, $\beta=0$) and the Schwarzschild BH ($a=0$, $\beta=0$). Conversely, on the right panel of figures \ref{fig1}, \ref{fig2}, and \ref{fig3}, for negative $\beta$, the energy flux is lower compared to the Kerr and Schwarzschild BHs. Furthermore, it is notable that as the $\beta$ value increases, the radius of the maximum flux gradually shifts towards lower $r/M$, indicating that most of the radiation originates from within the accretion disk. As shown in \cite{Collodel:2021gxu}, the energy flux of the disk around a scalarized Kerr BH is lower than that of a Kerr BH, meaning that the physical observable quantities of thin disk cannot distinguish between a scalarized Kerr BH and a BWBH with negative tidal charge.

\subsection{The radiation temperature}
Now we explore the radiation temperature $T(r)$ of disk. In the context of a steady-state thin disk model, accreting matter is typically assumed to be in thermal equilibrium, implying that the radiation emitted by the disk can be considered as perfect black-body radiation. The disk's radiation temperature $T(r)$ is related to the energy flux $F(r)$ through the Stefan-Boltzmann law, $F(r)=\sigma_{\rm{SB}}T^4(r)$, where $\sigma_{\rm{SB}}$ is the Stefan-Boltzmann constant.
%%%%%%%%%%%%%%%%%%%%%%
\begin{figure}
	\begin{minipage}{0.4\textwidth}
		\includegraphics[height=5cm,width=6cm]{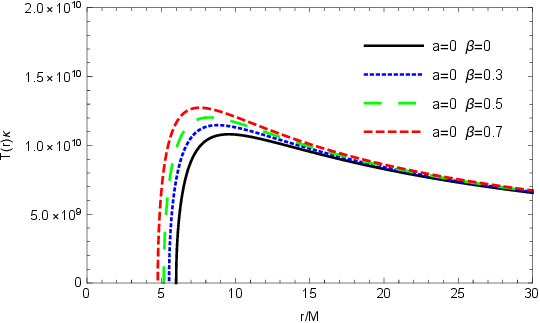}
	\end{minipage}%
	\begin{minipage}{0.4\textwidth}
		\includegraphics[height=5cm,width=6cm]{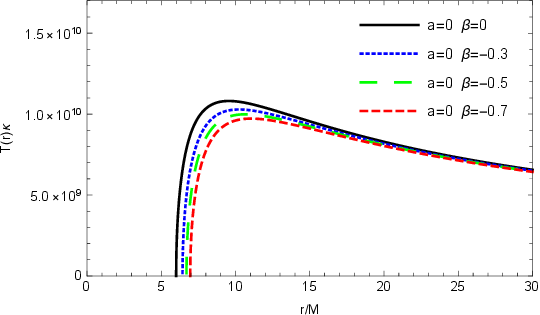}
	\end{minipage}
	\caption{ The radiation temperature $T(r)$ of a disk around a non-rotating $(a=0)$ BWBH with different values of $\beta$.}
	\label{fig4}
\end{figure}
%%%%%%%%%%%%%%%%%%%%%%%%%%%
\begin{figure}
	\begin{minipage}{0.4\textwidth}
		\includegraphics[height=5cm,width=7cm]{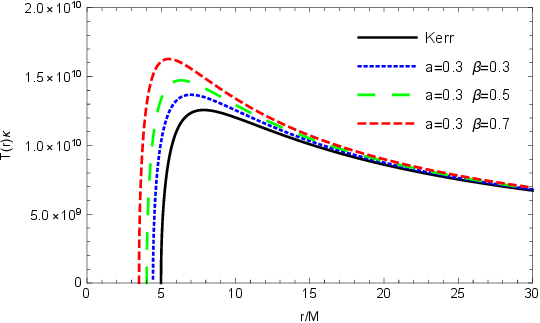}
	\end{minipage}%
	\begin{minipage}{0.4\textwidth}
		\includegraphics[height=5cm,width=7cm]{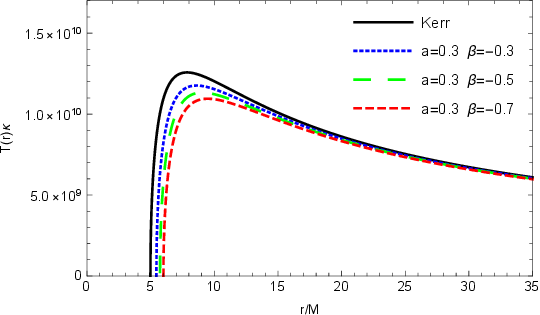}
	\end{minipage}
	\caption{ The radiation temperature $T(r)$ of a disk around a rotating $(a=0.3)$ BWBH with different values of $\beta$.}
	\label{fig5}
\end{figure}
%%%%%%%%%%%%%%%%%%%%%%%%%%
\begin{figure}
	\begin{minipage}{0.4\textwidth}
		\includegraphics[height=5cm,width=7cm]{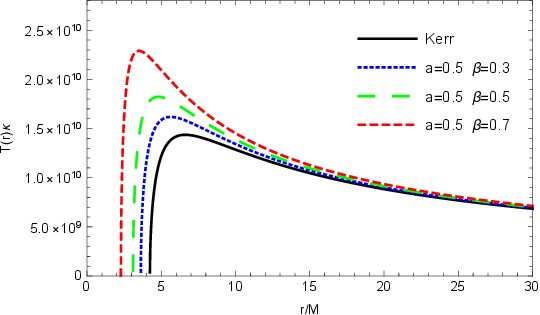}
	\end{minipage}%
	\begin{minipage}{0.4\textwidth}
		\includegraphics[height=5cm,width=7cm]{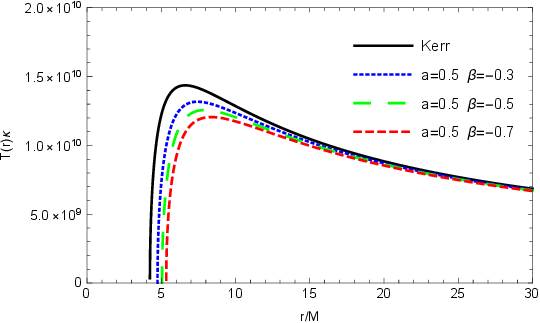}
	\end{minipage}
	\caption{ The radiation temperature $T(r)$ of a disk around a rotating $(a=0.5)$ BWBH with different values of $\beta$.}
	\label{fig6}
\end{figure}
%%%%%%%%%%%%%%%%

Figures \ref{fig4}, \ref{fig5} and \ref{fig6} depict the radiation temperature on the disk surface for different values of the parameters. Similar to case of the energy flux,
with increasing tidal charge $\beta$, the radiation temperature increases and the maximum values shift closer to the inner edge of the disk.
%Figures \ref{fig5} and \ref{fig6} show that as the rotation speed $a$ increases, the radiation temperature values increase more rapidly as $\beta$ increases.
On the left side of figures \ref{fig4}, \ref{fig5}, and \ref{fig6}, it can be observed that for positive $\beta$, the disk around a BWBH is hotter than that around a Kerr ($a\neq0$, $\beta=0$) or a Schwarzschild ($a=0$, $\beta=0$) BH. Conversely, on the right side of figures \ref{fig4}, \ref{fig5}, and \ref{fig6}, for negative $\beta$, the disk is cooler compared to the cases of Kerr and Schwarzschild BHs.

\subsection{The observed luminosity}
The luminosity $L(v)$ of a thin accretion disk surrounding a BH is inferred from the red-shifted black-body spectrum, as discussed in \cite{Torres:2002td}.
\begin{equation}\label{21}
L(v)=4\pi \mathrm{d}^{2}I(v)  =\frac{8\pi h\cos \gamma }{c^{2} } \int_{r_{\rm{i}} }^{r_{\rm{f}} } \int_{0}^{2\pi }\frac{v^{3}_{\rm{e}}r\mathrm{d}\phi \mathrm{d}r   }{e^{\frac{hv_{\rm{e}} }{k_{\rm{B}} T} }-1 },
\end{equation}
where $d$ is the distance from the center of the disk, $I(v)$ represents the heat flux radiated by the disk, $h$ is the Planck constant, $k_{\rm{B}}$ is the Boltzmann constant and $\gamma$ is the inclination of the disk (taking $\gamma$=0 here). The symbols $r_{\rm{f}}$ and $r_{\rm{i}}$ denote the outer and inner radii of the disk edge, respectively. We assumed that the flux on the disk surface diminishes at $r_{\rm{f}}$, hence we set $r_{\rm{i}}=r_{\rm{isco}}  $ and $r_{\rm{f}}\to \infty $. The emitted frequency is given by $v_{\rm{e}}=v(1+z)$. The redshift factor $z$ can be written as
\begin{equation}\label{22}
1+z=\frac{1+\Omega r\sin \phi \sin \gamma  }{\sqrt{-g_{tt} -2\Omega g_{t\phi }-\Omega ^{2} g_{\phi \phi }  } }.
\end{equation}
Here the bending effect of light is ignored\cite{Luminet:1979nyg,Bhattacharyya:2000kt}.
%%%%%%%%%%%%%%%%%%%%%%%
\begin{figure}
	\begin{minipage}{0.4\textwidth}
		\includegraphics[height=5cm,width=7cm]{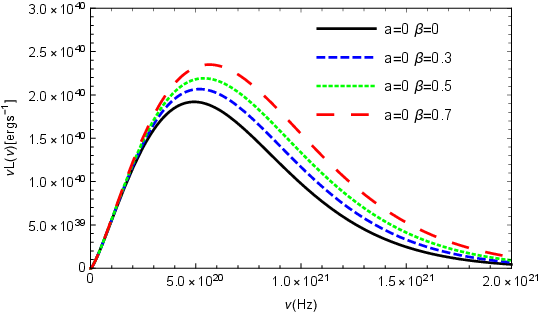}
	\end{minipage}%
	\begin{minipage}{0.4\textwidth}
		\includegraphics[height=5cm,width=7cm]{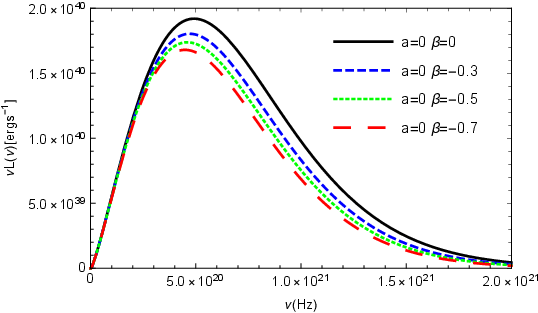}
	\end{minipage}
	\caption{The emission spectrum $vL(v)$ of disk around a non-rotating BWBH $(a=0)$ with different values of $\beta$. }
	\label{fig7}
\end{figure}
%%%%%%%%%%%%%%%%%%%%%%%%%
\begin{figure}
	\begin{minipage}{0.4\textwidth}
		\includegraphics[height=5cm,width=7cm]{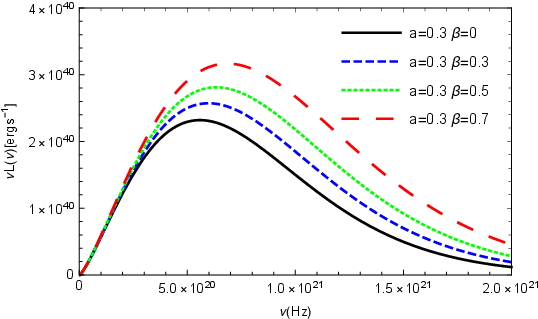}
	\end{minipage}%
	\begin{minipage}{0.4\textwidth}
		\includegraphics[height=5cm,width=7cm]{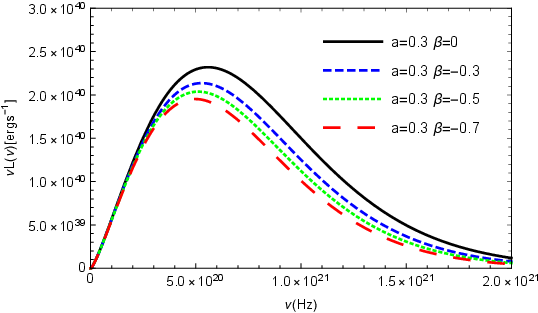}
	\end{minipage}
	\caption{The emission spectrum $vL(v)$ of disk around a rotating BWBH $(a=0.3)$ with different values of $\beta$.}
	\label{fig8}
\end{figure}

As shown in figures \ref{fig7} and \ref{fig8}, we can clearly see the spectral energy distributions of accretion disks around BWBHs. Similar to the energy flux and radiation temperature, as seen from the left side of figures \ref{fig7} and \ref{fig8}, for positive tidal charge, the spectral energy of disk around a BWBH is higher (more luminous) than that around a Kerr ($a\neq0$, $\beta=0$) or a Schwarzschild BH ($a=0$, $\beta=0$). Conversely, for negative $\beta$, the spectral energy of disk around a BWBH is lower (less luminous), compared to the cases of Kerr and Schwarzschild BH.

\subsection{The radiative efficiency and some important values}
During mass accretion around a BH, radiation efficiency is a key feature that describes the ability of the central object to convert its resting mass into outward radiating energy. In general, this conversion efficiency can be expressed by measuring the ratio of two rates: the rate of radiation energy escaping from the disk surface to infinity and the rate of mass energy transfer of the central dense object during mass accretion \cite{Page:1974he,Novikov:1973kta}. If all the emitted photons can successfully escape to infinity, then the efficiency $\epsilon$ will be determined by the specific energy of the particle at the edge stable orbit $r_{\text{isco}}$ \cite{Page:1974he,Novikov:1973kta}.
\begin{equation}\label{23}
	\epsilon  =1-E_{\text{isco}}
\end{equation}
\begin{table}
	\centering
	\begin{tabular}{c c c c c c  c c }
		\hline
		\hline
		$a$& $\beta$ & $\bar{r}_{\rm{isco}}$ & $F_{\rm{max}}(\bar{r})$ergs$^{-1}$cm$^{-2}$ & $T_{\rm{max}}(\bar{r})$K & $\nu_{\rm{crit}}$Hz & $\nu L(\nu)_{\rm{max}}$ergs$^{-1}$  &  $\epsilon$\\
		\hline
		0  & $-0.5$&  $6.69$&  $5.64\times10^{35}$ & $9.99\times10^{9}$  &  $4.59\times10^{20}$  &  $1.74\times10^{40} $   &0.052\\
		\hline
		&  $-0.3$ &  $6.43$  &  $6.34\times10^{35}$  &  $1.03\times10^{10} $  & $4.71\times10^{20} $  &  $1.80\times10^{40} $  & 0.054\\
		\hline
		 &  $0$  &  $6 $  &  $7.74\times10^{35}$  &  $1.08\times10^{10}  $  & $4.92\times10^{20}$  & $1.92\times10^{40} $  & 0.057\\
		\hline
		 & $0.3$  & $5.52$  & $9.82\times10^{35} $ &  $1.15\times10^{10} $  &$5.17\times10^{20} $  &  $2.07\times10^{40} $ &0.062\\
		\hline
		  & $0.5$  &  $5.17$  &  $1.19\times10^{36} $&  $1.20\times10^{10}  $&  $5.38\times10^{20}  $  &  $2.19\times10^{40} $  &0.065\\
		\hline
		0.3 & $-0.5$  & $5.73$  &  $9.33\times10^{35}  $  &  $1.13\times10^{10}  $ &  $5.11\times10^{20} $ & $2.04\times10^{40} $  &0.061\\
		\hline
		 &$-0.3$  & $5.45$ & $1.09\times10^{36} $ & $1.18\times10^{10} $ & $5.28\times10^{20}  $& $2.14\times10^{40} $ &0.064\\
		\hline
		&$0$ &  $4.98$ &$1.42\times10^{36} $&$1.26\times10^{10} $&$5.58\times10^{20} $&$2.32\times10^{40} $&0.069\\
		\hline
		 &$0.3$  & $4.44$ & $1.99\times10^{36}$ &$1.37\times10^{10} $&$ 5.98\times10^{20} $&$2.57\times10^{40}$&0.077\\
		\hline
		&$0.5$&$4.02$&$2.67\times10^{36} $  &  $1.47\times10^{10}  $&$ 6.34\times10^{20} $&$2.81\times10^{40} $ &0.084\\
		\hline
		0.5 & $-0.5$  & $5.05$  &  $1.42\times10^{36}  $  &  $1.26\times10^{10}  $ &  $5.58\times10^{20} $ & $2.32\times10^{40} $  &0.069\\
		\hline
		 &$-0.3$  & $4.74$ & $1.71\times10^{36} $ & $1.38\times10^{10} $ & $5.80\times10^{20}  $& $2.46\times10^{40} $ &0.074\\
		\hline
		&$0$ &  $4.23$ &$2.41\times10^{36} $&$1.43\times10^{10} $&$6.21\times10^{20} $&$2.73\times10^{40} $&0.082\\
		\hline
		 &$0.3$  & $3.62$ & $3.89\times10^{36}$ &$1.62\times10^{10} $&$ 6.82\times10^{20} $&$3.16\times10^{40}$&0.095\\
		\hline
		&$0.5$&$3.09 $& $6.26\times10^{36} $  &  $1.80\times10^{10}  $&$ 7.43\times10^{20} $&$3.63\times10^{40} $ &0.110\\
		\hline
		\hline
	\end{tabular}
	\caption{The maximum values of the energy flux $F_{\rm{max}}(\bar{r})$, the disk temperature $T_{\rm{max}}(\bar{r})$, and $\nu L(\nu)_{\rm{max}}$ for a thin accretion disk surrounding a BWBH. The dimensionless ISCO radius, the cut-off frequency $\nu_{\rm{crit}}$, and the Novikov-Thorne efficiency $\epsilon$ are also presented.}\label{tab1}
\end{table}

In table \ref{tab1}, we present the maximum values of the energy flux $F_{\rm{max}}(\bar{r})$, the radiation temperature $T_{\rm{max}}(\bar{r})$, and $\nu L(\nu)_{\rm{max}}$ for a thin accretion disk around a BWBH, along with the dimensionless ISCO radius, the critical cutoff frequency $\nu$, and the Novikov-Thorne efficiency $\epsilon$. It is observed that these values increase with an increase in the tidal charge $\beta$. Specifically, for negative $\beta$, these values are smaller compared to the Schwarzschild and Kerr BH; conversely, for positive $\beta$, the trend is reversed. The ISCO radius of disk around a BWBH decreases with increasing of tidal charge $\beta$. Furthermore, it is clear that as the rotation parameter $a$ increases, the difference in the flux maximum between the rotating BWBH and the kerr BH $(a\neq0,\beta=0)$ also increases. For increasing tidal charge $\beta$ with fixed $a$, the radiative efficiency increases; and for increasing $a$ with fixed tidal charge, the radiative efficiency increases faster.

As shown above, we see that the tidal charge has a significant impact on the physical properties of thin accretion disk, which maybe tested along with the observations of the shadows of Sgr A* and M87*. In \cite{Vagnozzi:2022moj}, the tidal charge was constrained from the shadow of Sgr A* as: $-0.01\leq q (=\beta/4M^2)\leq 0.15$ at 1 $\sigma$ confidence level for a non-rotating brane world black hole. In \cite{Neves:2020doc}, the tidal charge was constrained from the shadow of M87* as: $\beta \leq 0.004M^2$ for a BWBH with a cosmological constant. In \cite{Amarilla:2011fx}, it was found that the effect of a positive $\beta$ is to decrease the size of the shadow, while negative values generate an enlargement of its size.

\section{Conclusions}	
This study employed the Novikov-Thorne model to investigate the characteristics of thin accretion disks surrounding BHs in the brane world scenario. The equation for the ISCO is solved numerically. The analysis revealed a decrease in the ISCO radius of thin disks around BWBHs as the tidal charge $\beta$ increases. The impact of $\beta$ on the energy flux, the radiation temperature, the luminosity spectrum, and the energy conversion efficiency of the thin disk is explored under specific values of the rotation parameter $a$. The findings indicate a positive correlation between $\beta$ and these quantities. Furthermore, it is observed that under identical rotation parameters and positive $\beta$, the ISCO radius of BWBHs is smaller than that of Kerr BHs. Conversely, when $\beta$ is negative, the ISCO radius of BWBHs surpasses that of Kerr BHs.

Notably, the observable characteristics of the thin disk, including the maximum values of $F_{\text{max}}(r)$, $T_{\text{max}}(r)$, and $vL(v)_{\text{max}}$, demonstrate that the thin accretion disk around a BWBH is hotter and brighter for positive $\beta$ compared to Schwarzschild and Kerr BH in GR. Conversely, it appears colder and dimmer for negative $\beta$.

Given the significant differences in energy flux, the radiation temperature, the luminosity spectrum, and the conversion efficiency between BWBHs and those in GR, these findings suggest the potential for distinguishing between standard GR and brane world gravity based on observable characteristics.

\bibliographystyle{ieeetr}%{elsarticle-num}
\bibliography{mo}
\end{document}